\documentclass[12pt]{article}

\usepackage{epsfig}
\newcommand{\IR}{\hbox{I\hskip -2pt R}}

\def\frl{Friedmann-Lema\^ \i tre}
\def\coord {coordinate}
\def\spt {space-time}
\def\dS{de~Sitter}
\def\ie {{i.e.}}
\def\eg {{e.g.}}
\def\mink {Minkowski}
\def\R{{\rm I\!R}}
\def\gr {general relativity}\def\guill {\textquotedblleft ~} \def\minks {Minkowski space-time} \def\argcosh {\mbox{argcosh}} \newcommand{\der}[2]{\frac{d#1}{d#2}}
\newcommand{\pder}[2]{\frac{\partial#1}{\partial#2}}

\title{Cosmological effects in the local static frame} 
\author{Michel Mizony$^1$ and  Marc Lachi{\`e}ze-Rey$^2$ \\  \normalsize (Michel.Mizony@univ-lyon1.fr,~marclr@cea.fr) \\ \normalsize 1. 
Institut Girard Desargues, CNRS UMR 5028, \\
\normalsize B\^atiment Jean Braconnier, D\'epartement de math\'ematiques,  Universit\'e Lyon I \\
 \normalsize 43, Boulevard du 11 Novembre 1918
 F-69622 Villeurbanne Cedex
  \\
\normalsize 2.  Service d'Astrophysique, CE  Saclay, 
91191 Gif sur Yvette Cedex, France \\
\normalsize and  
CNRS  FRE K25910 \guill  Plasmas, Gravitation et Cosmologie "  } 
 
\date{October 2004}

\begin{document}

\maketitle
\abstract{What is the influence of cosmology (the expansion law and its acceleration, the cosmological constant...) on the dynamics and optics  of  a local system like the solar system, a galaxy, a cluster, a supercluster...? 
The answer requires the solution of Einstein equation  with the local source, which tends towards the cosmological model at large distance. 
There is, in general, no analytic expression for the corresponding metric,  but we calculate here an  expansion in a small parameter, which allows to answer the question.
First, we derive a static expression for the pure cosmological (\frl) metric, whose validity, although local, extends in a very large neighborhood of the observer. This expression appears as the metric of an osculating de Sitter model. 

Then we propose an expansion of  the cosmological metric with a local source, which is valid  in a very large neighborhood of the local system. This allows to calculate exactly the (tiny) influence of cosmology on the dynamics of the solar
system: it results that, contrary to some claims, cosmological  effects 
   fail to account for the unexplained acceleration of the Pioneer probe 
by several  order of magnitudes. Our expression provide   estimations 
of the cosmological influence in the calculations of rotation or dispersion velocity curves in galaxies, clusters, and any type of 
cosmic structure, necessary for  precise   evaluations of dark matter and/or cosmic flows.  
The same metric can also be used to estimate the influence of cosmology on gravitational optics in the vicinity of such systems.
}
\newpage

\section{Introduction}
{\bf Motivation}

  Dynamical studies in  the universe are usually separated into 
cosmological problems and local problems. For the first ones,    the 
Fried\-mann-Lema\^\i tre (FL) metric $g_{FL}$ is  most  often used, which is an exact solution of the  the Einstein equations (that we consider  in this paper as including the cosmological constant). For \emph{local} problems, as far as the gravitational field is weak (see below), one applies a Newtonian metric $g_{Nw}$, which is only an approximate solution of the Einstein equations for weak (\guill Newtonian ") fields.

It is clear that the validity of Newtonian  metric does not  extend at cosmological distances from the source, even for weak fields; for instance, it would not account for the cosmological redshifts. It is
 a requirement of \gr ~to reduce to Newtonian physics  for \emph{very local} (in a sense to be specified
below) problems.
However, the question of   intermediate regimes remains open. The first 
  goal of this paper is    to specify the meaning of \guill intermediate 
"; and    to provide an approximation (since no exact solution exists) 
for it. This intends to answer the questions: \guill what is  the  
  influence of cosmology for a local system? ".

Although cosmological effects are negligible in the solar system (this will be shown quantitatively below), they may influence the dynamics in 
    galaxies, in the Local group,  around clusters and  superclusters...
  We intend here to  evaluate exactly    the influence of  cosmology 
(the Hubble constant, the cosmological constant...)   inside  such  
systems. For the Solar System for instance, it has been questionned if
  the unexplained acceleration of the Pionner probe (\cite{Anderson}, \cite{Anderson04}) could be due to some cosmological effect. We will give a clear and quantitative
(negative) answer by  showing that such effect  remain absolutely tiny.  
This study applies to evaluate the influence of cosmology on the   
dynamics in a galaxy, or in its neighborhood: for instance, what is the 
(weak) contribution of cosmology to the rotation curves ?    The 
cosmological influence increases with the size of the system. Thus, we 
may infer that it must really   be taken into account (at some order 
that we will  precise)  for the dynamical studies inside or around galaxy clusters and  superclusters, as well for the study of gravitational lensing involving these systems.

{\bf Small parameters}

To evaluate the  \guill   local "  character with respect to cosmology, we will define 
  a natural parameter   $\epsilon = r~H_0$, where $r$ approximates the size of the system (the distance to  the gravitational
  source), and $H_0$ is  the Hubble constant, whose inverse provides a cosmological distance scale. Fortunately, the other cosmological 
distance  $\Lambda ^{-1/2}$   is of same order (the spatial curvature 
radius is probably even greater), so that this unique approximation will be valid for all cosmological effects. The Newtonian metric is  a zero   order  term  in  $\epsilon$. This paper   explores the 
following  terms, and give their exact formulation up to second order, which is largely sufficient for all studies.

We emphasize that this development is distinct from the weak field approximation. In the latter,  the small parameter may be taken as the potential $\phi_{Nw}$ (we chose units such that $c=1$). The pure (FL) cosmological metric is at zero order. The Newtonian metric is at first order, the next order being  the post Newtonian one, that we do not 
consider   in this paper (although a generalization to post-Newtonian 
order is straightforward). Thus our results will   not apply in the 
neighborhood of  compact objects like neutron stars or black holes. In 
the solar system, they will be valid only   sufficiently far from the 
Sun... They apply without restrictions  to  galaxies, clusters, superclusters...

The form of the metric that we provide is perfectly  adapted to analyse and interpret astrophysical results of cosmological relevance in these systems, and also  to make the link with  the usual concepts of laboratory physics.

{\bf Local developments}

We consider a system with spherical symmetry,    isolated in the 
Universe, to approximate the Sun, a galaxy, a cluster... An exact  solution of Einstein equations  is given by the Schwarzschild metric $g_{Schw}$. Its  first order approximation in $\phi_{Nw}$ gives $g_{Nw}$. It is clear that none of them   takes  into 
account any   cosmological effect, like for instance the 
cosmological   redshifts:  $g_{Schw}$ is  the solution (of the Einstein 
equation)  for an isolated source in a \emph{static} universe without 
curvature,   not in the  expanding Universe. In fact, the 
solution has  for  limiting condition   a flat  (\mink) \spt ~at infinity.

We search the solution    $g$ to the same equation,  with the limiting 
condition that the metric identifies to  $g_{FL}$ at infinity :   $g_{Schw}$ is its  approximation at  zero order  in the parameter $\epsilon$.   
  Excepted in special cases,   $g$ cannot be obtained exactly. Our main result if to provide   a static approximation $G$ of $g$ valid in a specified range.

A  metric is independent of a choice of coordinates (a map). It is however conveniently expressed in a given map. One difficulty of the 
problem is to find a convenient map  to express   $g$ or $G$.   For a 
\emph{global} analysis, the map where  $g_{FL}$ takes the usual   Robertson - Walker form     appears clearly   the most convenient. 
This is not so for local studies. Thus, to facilitate local studies, we must first  chose a map which provides a static form for   the pure cosmological  metric $g_{FL}$. 
Static solutions of the einstein equations are known and classified \cite{Stephani} but none identifies with $g_{FL}$, excepted in the   particular case of de Sitter  \spt. Thus, we are led to find an map which provides approximatively (in a well defined sense) a static form of $g_{FL}$. We obtain this form    $G_{FL}$  in section \ref{general}. It is exact   at order $\epsilon^2$, which is      largely 
sufficient for most calculations: even if we know the   exact cosmological solution  $g_{FL}$, we rewrite it in its approximate   form $G_{FL}$ to prepare the following.

This allows us to  calculate (in   section \ref{metric})  with the same    approximation (at order~$\epsilon ^2$), a static form  $G$ of   $g$, the    solution of  Einstein  equations with the central source, and  with the limiting FrL conditions.  
We resume in  the table the constraints which apply to the metric G:\\
- it is static;  \\
-   $G  \approx g_{Nw}$ at   order  $\epsilon ^0$.\\
- $  G   \approx    G_{FL}$ at zero order in $\phi _{Nw}$ 
(\ie, when  the source is off).

\begin{tabular}{|cc|c|c|c|c|}
\hline
   order  & &$\epsilon^0$& $\epsilon^1$&$\epsilon^2$...  &  \\
\hline
zero: &$(\varphi_{Nw})^0$   &  $g_{FL,0}= \eta_{Mink}$  & $g_{FL,1}$& $g_{FL,2}$...& $g_{FL}$ \\
Newtonian:& $(\varphi_{Nw})^1$   &  $g_{Nw}$  &  &$G$&\\
exact  :    & &   $g_{Schw}$ &  &&$g$\\
\hline \end{tabular}\\
\\ Table 1. The different metrics at different orders with respect to the small parameters 
$\varphi_{Nw}$ and $\epsilon$. \\

Thus, $G$ is the correct solution to explore the 
dynamics of the system beyond the near environment of the source. \\

{\bf The geometrical Point of View}

In plane  geometry, the local   study of a curve involves, at first
order, its tangent and, at second order, its osculating circle. The situation is  analog for the \spt ~manifold $({\cal U},g)$.
Its  first order approximation,  the (Minkowskian) tangent space, 
completely   neglects  any curvature (gravitational) effects:  the 
local deformations due to local sources, as well as the  imprint of the cosmic curvature (the  accelerated cosmic  expansion).
Note that $g_{Schw}$ takes only the local one and $g_{FL}$  the cosmological  one.
It   is of special  importance to retain both, when one is interested 
to   structures with a large extension (\eg, clusters or superclusters), 
where expansion affects the dynamics. The approximate metric $G$   
obtained below  exactly provides the second order  approximation of $({\cal U},g)$ which takes all these effects into account.
  We claim that this is the best second order approximation for \spt, the analog of the osculating
circle to a curve.  As we will see, it appears to be of constant curvature and, thus, may be called  the  \textquotedblleft ~osculating de
Sitter " \spt.

{\bf  Definitions}

We recall some definitions.\\
An \emph{inertial} frame is defined so that   the time coordinate
coincides with the proper time of a free-falling  observer, implying \begin{equation}g_{00}=1,~~~g_{0i}=0.
\end{equation}
A locally \emph{inertial} system of
coordinates around an event $E$ is an inertial one such that the metric is the Minkowski metric at $E$. \cite{Weinberg} (p. 127)  underlines the  usefulness of this kind of frame to understood the stress-energy tensor: \guill  Note that $p$ and $\rho$
  are always defined as the pressure and energy density measured by an observer in a locally inertial frame that happens to be moving with the fluid at the instant of measurement, and are therefore scalars ".\\ The principle of equivalence guarantees the
existence of a local    \emph{inertial} form of any metric.

In a \emph{static} system of coordinates,  all coefficients of the metric are independent of the time coordinate.

\section{The static   expression  of the cosmological metric}

This section considers the   purely cosmological case: we describe  the Universe  by a  Friedmann-Lema\^ \i tre    
model
$({\cal U},g_{FL})$, \ie,     spatially homogeneous and  isotropic. 
Most often, $g_{FL}$ is expressed
in its   Robertson-Walker form
\begin{equation}\label{RW} 
  g_{FL}=~ds_{FL}^2=d\tau ^2-R^2~(\tau)~[dx^2+f_{k}^2(x)~d\omega ^2],
      \end{equation}where $f_{k}(x)=x, ~\sin (x),~\sinh (x)$ according 
to the sign    $k= 0,~1$ or $-1$ of the spatial curvature, and where  
$d\omega ^2$ is the element of spherical angle.

We intend to   give a  different expression  of \emph{the same} metric~$g_{FL}$, which is static.  Since we know that there is no exact analytical solution in general, we will express it as a Taylor development in the small quantity $\epsilon$.   
We do not demand that it is global, \ie, that its validity extends to whole \spt ~(there would be no solution), but  only to   a sufficiently  extended neighbourhood of the observer today, defined as the  event\\ 
$E_0=$(here, today)$    \equiv \{\tau =\tau
_o,\ x= \theta =\phi =0\}$.

It is convenient to start from  the    {\bf locally inertial} (non static) form 
relative to $E_0$, defined  after defining  $\rho:=R~(\tau _o)~x=R_o~x$, as \begin{equation}\label{inertial} g_{FL} =d\tau ^2 -\frac{R^2(\tau)}{R^2(\tau _o)}~\left(d\rho ^2+R^2(\tau _o)~f_{k}^2(\frac{\rho }{R(\tau _o)})~d\omega ^2\right)\ .\end{equation} The slight change with respect to (\ref{RW}) emphasizes that, rigorously, an inertial metric must have the
Minkowskian form, and   that, in  (\ref{RW}), $x$ is an
\emph{angular} coordinate.

\subsection{  The \dS ~case}

For pedagogical reasons, and because some exact solutions can be found, we first examine the case of the \dS ~Universe. As usual, we describe \dS ~space-time as  the hyperboloid ${\cal H}$  isometrically embedded in 5-dimensional \minks ~ $\R ^{1+4}$. It is invariant under the de
Sitter  group SO(1,4).   Its constant
curvature radius is $\lambda ^{-1}$.  Here, $3 \lambda ^2= \Lambda$, the true (constant) cosmological  constant.

  We start from the usual (RW) form of the expanding  (${k}=-1$) metric: 
\begin{equation}\label{RWdS}
g_{dS} = d{\tau}^{2}
- \frac{\sinh^{2}{\lambda \tau}}{{\lambda}^{2}}(d{\alpha}^{2} + \sinh^{2}{\alpha}~ d{\omega}^{2})\ .\end{equation} The scale factor is $\sinh (\lambda \tau)/\lambda$ and the Hubble parameter
$H(\tau)=\lambda ~\coth (\lambda \tau)$. We   define the generalized
density parameter (which, here, includes the only contribution of the cosmological
constant) at time $\tau$ as $\Omega  =\Omega (\tau) =\Omega _{\Lambda} (\tau)=\lambda ^{2}/H(\tau)^{2}$ (we recall that $\lambda$ is a constant).
  The   Einstein equations  leads to
\begin{equation}\label{dSparameters}
H^2(\tau)-\frac{\lambda ^2}{\sinh^{2}{\lambda \tau}}=\lambda ^2=H^2(\tau)\
\Omega_{\Lambda}(\tau) .
\end{equation}

To obtain a  static expression, we will make two successive changes  of variables. The  first one,  $(\tau,\alpha) \mapsto (\tau,r)$ with  $r \equiv \frac{\sinh{\lambda}\tau}{{\lambda}}~\sinh{\alpha}$,  gives \begin{equation} g_{dS}  = d{\tau}^{2} - \frac{[dr -r~H(\tau)~ d\tau ]^{2}
}{1+\frac{r^2 \lambda
^2}{\sinh^{2}{\lambda \tau}}}- r^{2}d{\omega}^{2}.\end{equation} We now search a new change of variables  $(\tau,r) \mapsto (t,r)$ which supresses the cross term, \ie, provides  a synchronous form of the metric. To do so, we set $\tau = h(t,r)$, which implies $d\tau = \pder{h}{t}  ~dt + h'(t,r)~dr$. This requires \begin{equation}
h'(t,r) \equiv \pder{h}{r} =\frac{-r~H}{1-r^2~H^2 +\frac{r^2 \lambda ^2}{\sinh^{2}{\lambda \tau}}},\end{equation} leading to the exact metric
      \begin{equation}
g_{dS} =\frac{(1-r^2H^2 +\frac{r^2 \lambda ^2}{\sinh^{2}{\lambda
\tau}})}{1+\frac{r^2 \lambda ^2}{\sinh^{2}{\lambda \tau}}}~\left(\pder{h}{t} \right)^2~dt^2
\end{equation}
$$-\frac{1}{1-r^2H^2 +\frac{r^2 \lambda ^2}{\sinh^{2}{\lambda \tau}}}\ dr^2\ -\ r^2\ d\omega ^2\ .$$

The latter can be simplified thanks to (\ref{dSparameters}): 
\begin{equation}
g_{dS} =\frac{1-r^2\lambda ^2}{1+\frac{r^2 \lambda ^2}{\sinh^{2}{\lambda \tau}}}~\left(\pder{h}{t} \right)^2~dt^2\ -\frac{1}{1-r^2\lambda ^2}\ dr^2\ -\ r^2\ d\omega ^2\ .
      \end{equation}

\subsubsection{The generalized   local Birkhoff form}

The integration of  the differential equation
      \begin{equation}
h'(t,r)=-\frac{rH}{1-r^2\lambda ^2}= \frac{-r~\lambda ~\coth[\lambda h(t,r)]}{1-r^2\lambda ^2},
	\end{equation} 
with the limiting   condition $h(t,0)=t$,  gives 

$h(t,r)=\frac{1}{\lambda}~\argcosh[\coth(\lambda
t)\sqrt{1-r^2\lambda ^2}]$, and   $ (\pder{h}{t} )^2=1+\frac{r^2 
\lambda ^2}{\sinh^{2}{\lambda \tau}}$.
Finally, we   obtain the desired change of
variables \begin{equation}(\tau,\alpha) \mapsto (t,r);~r=\sinh (\lambda \tau)~\sinh \alpha
/\lambda,~\tanh(\lambda t)=\cosh(\alpha)~\tanh(\lambda   \tau), 
\end{equation} and
the  {\sl local  static  Birkhoff form} of the (same)  metric \begin{equation}\label{BdS}  g_{dS}=(1-r^2\lambda ^2)~dt^2\ -\frac{1}{1-r^2\lambda ^2}\ dr^2\ -\
r^2\ d\omega ^2\ .    \end{equation}
This the exact de Sitter metric. Note however that  this expression  is only valid in a local map, namely inside the causal diamond of the observer (which is sufficient for any local problem).
In the  next section, we will apply a similar procedure  to the general FrL model. This will lead to a comparable  form, although   only exact at order
$(H_0~r) ^2$.

Geometrically, this may be seen as   a projection, in  $\IR ^{1+4}$, of 
the hyperboloid.
This projection,  orthogonal to  $e_1$, $e_2$ et $e_3$, with  center 0 in the  plane {$e_0$, $e_4$}, maps ${\cal H}$ to the \textquotedblleft ~hyperbolic nap ", which is a cylindric hypersurface with hyperbolic section.

We recall that the (local) Birkhoff theorem states that, for any metric with spherical symmetry, in a vacuum universe, there is a local coordinates system $(\tau,\rho,\omega)$ (corresponding to a peculiar
observer)  such that the metric coefficients do not depend on time. In \dS ~universe, the theorem leads to the metric above.
For such a form of the metric,   calculating the geodesics is  
particularly simple.

    \subsection{The general Friedmann-Lema\^ \i tre model} \label{general} 

    We now apply the same approach to the general FrL model, characterized by a scale factor 
$R(\tau)$,   $\tau$ being the cosmic time.  We have the Hubble 
parameter   $H(\tau) \equiv \dot
{R}(\tau)/R(\tau)$, the deceleration parameter $q(\tau) \equiv  -\frac{ \ddot {R}~R}{\dot{R} ^2}$, that we assume negative (corresponding to an 
accelerating expansion; this is essential for the  following calculations), and the      Einstein  equation   
\begin{equation}\label{Einstein}
H(\tau)^2+  \frac{k}{R(\tau)^2}=\frac{8\pi G\rho(\tau) }{3} \equiv \Omega (\tau) ~H ^2(\tau).\end{equation} We have  included in   $\Omega $  the contributions of matter and cosmological 
constant (or exotic energy). {\sl Present}  values of these quantities 
are   written with a zero index : $ \Omega (\tau _0) = \Omega _0 $, 
etc. Recent  observations favour $ \Omega _0 \approx 1,~q _0\leq -0.55$. 

We start
from the usual form  (\ref{inertial}) of the metric,  which is expressed in the
  locally inertial frame  for the event   $E_0 = $ (today, here).  The  
first change of  variables $(\tau,\rho) \mapsto   (\tau,r \equiv R(\tau
)~f_{k}[\frac{\rho }{R(\tau _o)}])$ leads to the form \begin{equation}
g_{FL}=d\tau ^2 -\frac
{[dr- r~H(\tau)~d\tau ]^2}{1 + [1-\Omega (\tau)]~H^2(\tau)~r^2} ~ \ -\ r^2\ d\omega ^2. \end{equation} A comoving galaxy is defined by $d\rho =0$, which implies $dr= r~H(\tau)~d\tau $. This  formulation of the Hubble law  remains exact at any time.

  To continue, we define like above   a new variable $t$ by  $\tau 
=h(t,r)$ and  eliminate  the cross term. This   requires   
\begin{equation}
h'(t,r)\equiv \pder{h}{r}=\frac{-r ~H[h(t,r)]}{1-   \Omega 
(\tau)~H^2(\tau)~r^2}, \end{equation}which  allows, using 
(\ref{Einstein}),   to write \begin{equation}\label{GB}
g_{FL} =\frac{1-\Omega  ~H^2 ~r^2}{1 + (1-\Omega  )~H^2 ~r^2}~\left(\pder{h}{t} \right)^2~dt ^2\ - \frac{1}{1-\Omega ~H^2~r^2}\ dr^2\ -\ r^2\ d\omega
^2  ,  \end{equation} where   $H$ and $\Omega$  are to  be seen as 
functions   $ H[h(t,r)],~ \Omega [h(t,r)]$.
We call (\ref{GB}) the \emph{generalized    Birkhoff  form}
of the model. It would be convenient to precise the change of variables $\tau = h(t,r)$, by solving  the differential  equation $$h'(t,r)=-\frac{rH[h(t,r)]}{1-\Omega [h(t,r)]~H ^2[h(t,r)]~r^2},\ h(t,0)=t.$$  This  would however require the explicit knowledge of $R(\tau)$, which  is not possible analytically, in general.
However, we are interested in \emph{local} solutions: we  obtain a development as
\begin{equation}
h(t,r)=t-\frac{H(t)}{2H_0^2}~(H_0~r)^2-\frac{H^3(t)}{H_0^4}~\frac{
[2\Omega (t)+q(t)+1]}{8}~(H_0~r)^4+ {\cal O}(5), \end{equation}where ${\cal O}(5)$ means terms of order at least 5 in $(H_0~r)$.
Reporting in~(\ref{GB}) above
 leads to  the following  local development form of the metric, at the vicinity of the event $E_0 \equiv (t=t_0, r=0)$:
\begin{equation}\label{AGB0}
\begin{array}{r}
g_{FL} = \left(1+ q(t) ~H^2(t)  \ r^2+[\Omega (t)+q (t)]~{\cal{O}}( 4)\right)~dt ^2\\ 
\ \ \ \ \ \ \ \ \ \ \ \ \ - \frac{dr^2}{1-\Omega (t)~H^2(t) \ r^2} -\ r^2\ d\omega ^2\ , 
\end{array}    
\end{equation}
where     $  H(t), \Omega (t),...$ are functions of $t$ only.

This form (\ref{AGB0})   is {\sl non static}, since   
$\Omega  ~H^2$ depends on time,
excepted for the  de Sitter models where $\Omega  ~H^2= -q~H^2=\lambda ^2$. To obtain the static approximation, we have to 
develop all quantities   with respect to $\delta t=t-t_0$, the present 
time.  For instance,  $H(t) =H_0~[1+(q-1)~H_0 ~\delta t+O(H_0 ~\delta t)^2]$.
However,   all time-dependent terms occur in  factor of $r^2$ (or 
higher order terms). This allows, at second order, to  replace all time-dependent quantities by their present values, for instance $H(t)$ by $H_0$, and similarly for $\Omega _0$ and $q _0$.
This leads to
 the approximate form
\begin{equation}\label{AGB}
G_{FL} = [ 1+ q_0~(H_0   \ r)^2]~dt ^2\ - [1+\Omega _0~(H_0~r)^2]~dr^2 -\ r^2\ d\omega ^2.\end{equation}

\begin{itemize}
\item  This form   $G_{FL}$ is static and we call it   the {\sl local static osculating metric}.
  \item  It  implies that the global (cosmological)   \spt ~curvature imprints 
  weak,  but non necessarily  negligible, effects  in the local dynamics,   through
the three   cosmological parameters $H_0 ,q_0 , \Omega _0$.
 \item Locally, and  for weak fields, the   variable $r$ has the   
meaning of a (spatial) curvature
radius, and thus of a   distance.
\item The  expression (\ref{AGB}) is at order two. 
However, it  identifies with the exact  static form  (\ref{BdS})  for the   de 
Sitter  model. This makes appear the latter as the   osculating \spt ~to     the \frl ~model, its second 
order approximation (the first order one     being the tangent Minkowskian 
space).
(We recall that, for an accelerating model of the Universe,  $q_0$ is negative.)
\item 
In the case of  a de Sitter model,   $\Omega  ~H^2= \lambda ^2=Cte$.
Then (\ref{AGB})  becomes an  exact expression, which  identifies to     (\ref{BdS}).
\item 
Since data indicate $\Omega_o   \approx 1, ~H_0 ^{-1} \approx  4.3 $~Gpc, the  metric above  gives a very good  precision $\approx (r/4  ~Gpc)^3$.
\end{itemize}

    \subsection{Geodesics and redshifts}

We derive in appendix A   (\ref{geod2}) the  geodesic equation, at 
second order, for a static metric.  The explicit form of the metric (\ref{AGB})
gives (at second order ) 
  \begin{equation}
  \gamma =H_0 ^2 \ r   ~[q_0~  + V_i^2~( \Omega 
_0-2q_0)].\end{equation}
This is the  radial   three-acceleration   of a particle in free fall, expressed in the local frame. It is  of pure 
cosmological origin.
Note that  second order terms are zero.

A \guill comoving " galaxy is defined by $\rho=\rho _0=C^{te}$. Writing this relation with the new coordinates, we obtain  the Hubble law $V = H~r$, valid at second order.

\section{The local system}

\subsection{The metric}\label{metric}

The previous   calculations provided   a static  approximation  of 
the cosmological metric $g_{FL}$, exact at second order in $H_0~r$ around a local observer. 
This  was only a first step to deal with  the  more interesting problem of a massive system in the cosmological context.

The problem is to find the Universe metric $G$, which  is the approximation at first order in $\phi _{Nw}$, and at second order in $H_0 r$, of  the exact  solution of Einstein equations $g$. At   the lowest (zero) order  in $H_0~r$, it describes   Newtonian  physics.  At zero order in $\phi _{Nw}$  (or, equivalently, when 
we put $M=0$), $G$  is equivalent to     $g_{FL}$, up to second  order 
in  $H_0~r$.

In the general case, it is not possible to find such a metric analytically. This is possible in the de Sitter case, as it appears 
below. However,   local studies do  no require an exact  knowledge of  
this metric: a development at the lowest orders in $H_0~r$ is sufficient. What order is needed, and what is the right development ? 
We answer these two questions below.

We assume a local mass overdensity  $M$ with spherical symmetry,    
isolated in cosmological space, intended to represent the Sun, a galaxy or a cluster, taken as the origin of the coordinates.
   The gravitational   field is assumed weak and all calculations are at 
first (Newtonian) order only. This excludes the account of any post-Newtonian effect.

We adopt the  coordinate system which generalizes that of 
equ.(\ref{AGB}). The solution of the Einstein equation (with a source) is given, at  second order in $H_0~r$, by  MAPLE as 
\begin{equation}
g= \left( 1-2\,{\frac {M}{r}}+q_0~(H_0 r)^{2} +{\cal O} (\epsilon ^3)\right) {{\it dt}}^{2}\end{equation} $$-{{ \it dr}}^{2} \left( 1-2\,{\frac {M}{r}}-\Omega _0\,(H_0 r)^{2}
  +{\cal O} (\epsilon ^3)\right) ^{-1}-{r}^{2}{{\it d{\omega}}}^{2},$$
which  reduce to   the cosmological form (\ref{AGB0})   when no mass is 
present.

The static approximation
\begin{equation}
\label{dsU2} g\approx G  \equiv  \left( 1-2\,{\frac {M}{r}}+q_0~(H_0 r)^{2}  \right) {{\it dt}}^{2}-{{ \it dr}}^{2} \left( 1-2\,{\frac {M}{r}}-\Omega _0\,(H_0 r)^{2}\right) ^{-1}
\end{equation}
$$-{r}^{2}{{\it d{\omega}}}^{2}$$  is at order 2.  It guarantees that cosmological effects are correctly taken into account at this order above, and that local effects are treated at the Newtonian level.
This formula is exact when the cosmological background is \dS ~(the well-known Schwarzschild-\dS ~metric; see, e. g.,  \cite{Rindler}).

This is to be contrasted with the pure Newtonian approximation:
\begin{equation}
\label{dsN} g_{Nw}=\left( 1-2\,{\frac {M}{r}}\right) {{\it dt}}^{2}-{{ \it dr}}^{2} \left( 1-2\,{\frac {M}{r}}
  \right) ^{-1}-{r}^{2}{{\it d{\omega}}}^{2},\end{equation} which is at the same order, but does not take any cosmological effect into account.
If we want  to  compare the two forms, we must be carefull   not to 
confuse real effects with  coordinate effects. This requires to consider only covariant quantities, which may however  involve the observer.

\subsection{The attraction radius}

A first application is the   definition of  the concept of 
\emph{attraction radius}   by a local overdensity in a non empty 
environment   \cite{Souriau},\cite{Mizony}:  playing the role of a  kind of 
cosmological tidal radius, it is defined  as the distance $r_M$ such 
that the Newtonian    contribution of  the overdensity
becomes equal and opposite to that   of cosmological origin: $r_M^{3} =
\frac{M}{{q_o~H_0}^{2}}$. This gives a rough estimate of the ratio of the influences of the  two contributions which goes like  local / cosmological  $\approx  (r_M/r)^{3}$.

The attraction radius due to the gravitational influence of the Sun 
amounts  to $ \approx 240 $~pc. Thus, cosmological effects are expected of the order  $   (r/240~\mbox{pc})^{3}$, much  smaller than the Newtonian ones (see below).
For a typical galaxy, $r_M \approx 1$~Mpc. For a  cluster like Virgo, $r_M \approx 20$~Mpc  indicates that cosmological effects may become non completely negligible; and {\sl a fortiori} for superclusters. 

\subsection{Dynamics in the local system} \subsubsection{Radial or tangential  free fall}

  The geodesic equations in a   static  metric are  derived in  Appendix 
A. Replacing the metric coefficients by their values (\ref{dsU2}), we obtain the radial geodesic equation $$r(t)=r_o + v_o (t-t_o) + 
\frac{\gamma (r_o)}{2}(t-t_o)^2+ ...,$$  with   
\begin{equation}\label{gamma}
\gamma
(r)=-\frac{M}{r^2}-q_0~H_0^2r+(2q_0~+\Omega_0)H_0^2~M\end{equation}
$$+\left(\frac{3M}{r^2}+(2q_0-\Omega_0~)~H_0~^2(r+M)\right)V^2(r), $$
  to be compared to
$$ \gamma _{Nw}=
  -\frac{M}{r^2}
   +(\frac{3M}{r^2})V^2(r). $$

   For tangential free fall,  the application of the formulae above to
(\ref{acctan}) gives
\begin{equation}\label{gammatheta}
r~V_{\theta} ^2=\gamma _{\theta} =\frac{M}{r^2}+q_0~H_0^2~ r,
\end{equation} 
\ie,  a contribution similar to the radial case.

\subsubsection{Observations}

  The motion of 
  a probe in the Solar system,   a star in a galaxy,   a  galaxy in a cluster, ...    is not directly observable. The only information available to the terrestrial observer   is the  redshift, that he can monitor as a function of its proper time. Such dynamical analyses are at the basis of the Pioneer effect, and of the estimations of dark matter in the astronomical structures. An exact estimation of the cosmological effects appears therefore  very important.
  
 The most direct effect is the addition  of  a cosmological component  to the Newtonian redshift (\ie, velocity, through Doppler effect). This may affect, for instance, the
     dark matter estimations from rotation or dispersion curves. Also,    an additional term  of cosmological origin is added  in the derivative of $z$ (w.r.t. observer proper time). Precisely, the Pioneer effect appears   under the form of 
such  an additional term, with  the dimension of an acceleration.  However, the calculation below clearly shows that, contrary to recent claims, it is definitely not of cosmological origin.

{\bf The Pioneer effect is not cosmological}

The inertial probe emits a light-ray     at $(t_s,r_s)$. It  is 
received by the observer at $(t=T,r=0)$. Relativistic optics (equ.\ref{tT})    relates $T$ (the time when the observer receives the light) to $t_s$. The observer monitors the redshift as a function $z(T)$ of his proper time, calculated in Appendix~B. He may calculate the derivative $a \equiv \der{z}{T}$.

We recall that the Pioneer effect \cite{Anderson},\cite{Anderson04}  as 
a time variation of the observed redshift of the   Pioneer probe, 
reported as an anomalous acceleration\\ $a_{Pioneer, observed} = (8.74 \pm 1.33) ~ 10^{-10} $~ m~s$^{-2}$\\  towards the sun, at a distance 20
  a.u.~$\approx 10^{-5} $ pc $\approx 4.5~ 10^{16}$m from it.

We  applied the calculations of appendix B  to the Pioneer probe,    a source in radial  free fall in the Solar System. This  gives a cosmological contribution $a_{Pioneer, cosmological }=q_0 H_0^2 r^2$. \\ Numerical estimations (with $h_0=.7$) give $5q_0~ 10^{-20} $ m~sec$^{-2}$, about $10^{10}$  orders of magnitude smaller than the observed effect.

\subsubsection{Cosmological  effects in extragalactic astronomy}

 Evaluations of dark matter in  galaxies or clusters
result  from redshift measurements of particles   (stars, gas molecules, galaxies) in free fall in the gravitational  potential.
The  formula (\ref{gammatheta}) is directly applicable to evaluate   the contribution 
of cosmology.

At the periphery of a galaxy (30 kpc), the cosmological acceleration reaches $a \approx 4\ 10^{-15}$m~sec$^{-2}$. This   corresponds  to a  velocity component  of about 2 km sec$^{-1}$, to be  compared  to a  typical Newtonian  rotation velocity  of several  100 km sec$^{-1}$.

At the periphery of a cluster of galaxies (5 Mpc), the  cosmological acceleration reaches $a \approx 6~10^{-13}$m~sec$^{-2}$. The additional  velocity,  about 300 km sec$^{-1}$, may be non negligible in precise estimations of dark matter. Its contribution increases with the size of the system, and thus becomes certainly important for estimations of  deceleration or acceleration beyond the   cluster size. In particular, the   metric $G$ is   perfectly adapted to explore the 
large scale velocity fields (like the Virgocentric flow), where cosmological and local  effects are of the same orders of magnitude: it   appears much  more convenient  than the Tolman-Bondi expression.  Such analyses are in progress.

In these situations, cosmology  affects not only the dynamics, but also the   gravitational optics (lensing of light rays). The importance of the latter, as astrophysical and cosmological tools, justifies 
the precise estimation of these effects thanks to the metric $G$.  Those will be estimated in a forthcoming paper.

\section{Conclusion }

The   metric of  a given \spt ~Êcan be written in    multiple forms. Each one may offer a peculiar geometrical and/or physical interest. Each one has its own    validity range,  in general lower than the whole  \spt ~manifold, what is well shown by the   isometric embedding in   $\IR ^5$  \cite{lach},\cite{Mizony}. 

Here we have provided a local static form (\ref{AGB}) for the pure cosmological metric. Although its use may be of high convenience   for a   local cosmological  study, its interest remains  rather  academic, since an exact (although non static) form is available. However, this emphasizes the fact that 
the local cosmology (in a wide sense) is that of the \dS ~\spt, which appears as the osculating \spt ~to any FrL cosmology.

Addressing the question of a   local overdensity in the   cosmological context, we have found a static solution  (\ref{dsU2}), which is exact at order $(H_0 r)^2$, largely sufficient for any  study in  the vicinity of  a local overdensity in the Universe. This clearly expresses the  influence of cosmology in such situations, for gravitational dynamics and optics.

We have evaluated precisely   the cosmological effects in the solar system. They appear negligible, which implies     unambiguously that   the unexplained Pioneer acceleration is not of cosmological origin. We have estimated   such  effects for  dynamical studies  of  galaxies, clusters, superclusters..., and in particular how they may affect dark matter estimations. 

Our  proposed metric form $G$  offers a perfect framework to study any situation where cosmological effects are not negligible compared to the local ones. This is the case for 
large scale velocity fields  like the local Virgocentric flow.
This is  also the case for  gravitational optics at the  periphery of large scale mass condensations.  This  powerful tool, for estimations of [dark] massive matter, cannot be used without correct estimations  of the cosmological effects, which   may become non negligible (work   in progress).

The metric $G$   offers  an unique framework  for the exploration of the extragalactic  Universe, where cosmological  effects superpose to the local ones.

\section{Appendix A: Geodesic equation}

  Inertial motion is given by  the geodesic equations. We give a complete derivation, which allows an adaptation to more complicate situations.  We derive the geodesic equations  as   the       Euler-Lagrange equations for   the Lagrangian ${\cal L}=1/2~g_{\mu \nu}~v^{ \mu}~v^{ 
\nu}$, with $v^{ \nu} \equiv \der{x^{ \nu}}{\lambda}$, and $\lambda$ is a time-parameter: \begin{equation} 
\frac{1}{2}~\pder{g_{\mu \nu}}{x^{\rho}}~v^{ \mu}~v^{ \nu}=\der{(g_{\mu \rho}~v^{ \mu})}{\lambda}.\end{equation} 
Here, the coefficients of the metric reduce to the diagonal ones, and 
depend on the coordinate $r$  only (static form).   The first equation 
($\rho=t$) implies that $v_t \equiv v^t~g_{tt}$ is a constant of motion. This is valid with any choice of the  parameter $\lambda$. In particular, for $ \lambda= \tau$ (the proper time), this equation gives $u_\tau=E=Cte$, which expresses  energy conservation along the motion.

{\bf Radial motion}

For  a radial timelike geodesics, the second one ($\rho=r$) is equivalent to the unitarity condition for the four velocity $u$. In our case, it is more convenient  to work with the latter, namely\begin{equation} g_{tt}~(u^t)^2+g_{rr}~(u^r)^2= (u_t)^2/g_{tt}~+g_{rr}~(u^r)^2=1.
\end{equation}
Defining a 3-velocity $V \equiv \der{r}{t}=\frac{u^r}{u^t}$, this can be rewritten $g_{tt}+ g_{rr}~V^2=( \frac{g_{tt}}{u_t})^2=( \frac{g_{tt}}{E})^2$, that we simply derive wrt $t$. This defines a 3-acceleration $\gamma \equiv \der{^2r}{t^2}$, and we obtain $$\gamma =\left( \frac{2g_{tt}~g'_{tt}}{u_t^2}-g'_{rr}~V^2-g'_{tt} \right)/(2 g_{rr}),$$ where the prime means the derivative wrt $r$. Taking into account initial conditions, \begin{equation}\label{init} g_{rr}(r_i)~V_i^2=g_{tt}^2(r_i)/(u_t)^2-g_{tt}(r_i),
  \end{equation} and developing, we obtain    the radial geodesic 
equation at second order,\begin{equation} \label{geod2} r(t)=r_i +V_i (t-t_i) + \frac{\gamma (r_i)}{2}(t-t_i)^2+ ...,\end{equation}
  with
\begin{equation}\label{acc0}
\gamma(r_i)=\left(\frac{1}{2g_{rr}}\frac{dg_{tt}}{dr}+\frac{V_i^2
}{g_{tt}}\frac{dg_{tt}}{dr}-\frac{V_i^2
}{2g_{rr}}\frac{dg_{rr}}{dr}\right)(r_i)\end{equation}
appearing as the radial three-acceleration.

{\bf Tangential motion}

Tangential motion is defined by $u^r=0$. The geodesic equations imply that all components of the velocity remain constant. Moreover, the second equation reads $$g'_{tt}~u^t~u^t+g'_{ \theta \theta }~u^{\theta}~u^{\theta}.$$ Joined to the unitarity  of the velocity, and defining $$V_{\theta} \equiv \der{\theta}{t}=\frac{~u^{\theta}~}{~u^{t}},$$ this gives \begin{equation} \label{acctan} r  ~V^2_{\theta} = \frac{g'_{tt}}{2}, \end{equation} where the RHS appears as the centripetal acceleration.

Calculating the redshift as a Doppler one, we have $z \approx V \approx \frac{M}{r}+q ~H^2~r^2$. This means that we have an additional term of cosmological origin when we measure a rotation curve in a galaxy, for instance. A numerical estimation  gives the  additional cosmic contribution as $\Delta V^2 =q H^2 r^2 \approx 4q_0  (km/s)^2$, for r=30 kpc.

\section{Appendix B: redshift in a static metric}

We assume a static metric   $ds^2=A^2(r)~dt^2-B^2(r)~dr^2 $.
We calculate the redshift (seen by the observer at the origin of \coord s, $r=0$) emitted by a (moving)  source in radial motion, $r=f(t)$, thus with a radial 3-velocity $V_s \equiv df/dt$. We assume the light 
ray also radial. Calculations are   at second order in $H_0 r $.

The light ray, emitted by the   source at  $(t_s,r_s)$, reaches the 
observer at   $(T,~r=0)$:
$$  T -t_s=\int _{0}^{r_s} \frac{B(r)~~dr}{A(r)}.$$

A short delay after, the source has moved to   $(t_s+\delta t_s, 
r_s+\delta r_s)$, and emits a second light-ray, which reaches  the 
observer at   $(T +\delta T ,r=0)$:
  \begin{equation}
\label{tT}
    T -t_s+  \delta T -\delta t_s=\int _{0}^{r_s+ \delta r_s}B(r)/A(r)~dr.
\end{equation}

Subtraction gives
$$\delta T -\delta t_s=\int _{r_s}^{r_s+\delta r_s}~\frac{B(r)}{A(r)}~dr \approx \frac{B(r_s)}{A(r_s)} ~\delta r_s.$$ Considering the two rays as two light fronts, as usual, we obtain the two periods of emission and reception:\\
$T_{emission}=[A(r_s)^2 ~\delta t_s^2-B(r_s)^2 ~ \delta r_s^2]^{1/2}$ (in  proper time of the source),\\ $T_{reception}=\delta T $, since $T$ measures the proper time of the observer. Thus (definition), the
  redshift: $1+z \equiv \frac{T_{reception}}{T_{emission}}=
\frac{ \delta T_1 }{[A(r_s)^2 ~\delta t_s^2-B(r_s)^2 ~ \delta r_s^2]^{1/2}} $.
Combining, with $\delta r_s=V_s~\delta t_s$,  some algebra leads to \begin{equation}
1+z=\frac{ B_s~V_s~+A_s }{A_s~[A_s^2 ~-B_s^2
~V_s^2~]^{1/2}}=\frac{1}{A_s} \sqrt{\frac{ [B_s~V_s~+A_s] }{ ~[A_s ~-B_s ~V_s]}}.
\end{equation}

Writing for simplification, $A_s=1+a$ and $B_s= 1+b$, with $a$ and $b$ second order quantities, we obtain the development \begin{equation} \label{redshift}
1+z  \approx  \left[  1-a+      V_s \frac{b-a}{1-V_s^2} 
\right]~\sqrt{\frac{ 1+V_s  }{ 1 -V_s}}~.\end{equation} When the source  is non relativistic ($V_s <<1$), this reduces to
  $$ z \approx V_s~(1+b-a) -a.$$ All subsequent calculations will be at first order in $V_s$.

Specifying the metric as in the text, this becomes   \begin{equation}
\label{lastz}
  z \approx V_s~[1+\frac{2M}{r}+(\Omega  -q)~\frac{H^2~r^2}{2}] 
+\frac{M}{r}-q~\frac{H^2~r^2}{2}.
\end{equation}  This is the redshift measured by the observer, that he 
can   register  as a function of his  proper time $T_1$. Calculating 
the derivative,    he will  read a three-acceleration $a \equiv 
\der{z}{T_1}=\der{z}{t_s}/~\der{T_1}{t_s}$, where the last derivative is calculated from (\ref{tT}) as $1+\frac{V_s ~B(r_s)}{A(r_s)} \approx 
1+ V_s (1-a+b)$.

  {\bf Radial inertial source}

  To calculate $\der{z}{t_s}$,  we assume the source in radial inertial 
motion
   (\ref{geod2}). From (\ref{lastz}),
$$ \der{z}{t_s}= \gamma (r_s)~[1+2\frac{M}{r}+(\Omega  
-q)~H^2~\frac{r^2}{2}]
-\frac{V_s}{r}~( \frac{M}{r}+ q~H^2~r^2),$$
where we have neglected quadratic terms in the velocity. Inserting 
$\gamma (r_s) \approx -a'~(1-2b)$, and $a'=\frac{M}{r^2}+q~H^2 r$, we 
obtain (at second order)
$$ \der{z}{t_s} \approx  - (\frac{M}{r^2}+q~H^2 r):$$
   we observe a   cosmic three-acceleration
  $-q~H^2 r$ in addition to the Newtonian one.

\end{document}